\documentclass[pra,superscriptaddress,reprint]{revtex4-1}
\usepackage[utf8]{inputenc}
\usepackage[fleqn]{amsmath} 
\usepackage{amsfonts}
\usepackage{amssymb}
\usepackage{graphicx}
\usepackage{epstopdf}
\usepackage[subrefformat=parens,labelformat=parens]{subfig}
\usepackage[T1]{fontenc} 
\usepackage{url}
\usepackage[hidelinks]{hyperref} 
\usepackage[toc,page,title,titletoc,header]{appendix} 

\begin{document}
\author{S. J. Cooper}
\email{scooper@star.sr.bham.ac.uk}
\author{C. J. Collins}
\author{A. C. Green}
\author{D. Hoyland}
\author{C. C. Speake}
\author{A. Freise}
\author{C. M. Mow-Lowry}
\affiliation{University of Birmingham, Birmingham, B15 2TT, UK}
\title{A compact, large-range interferometer for precision measurement and inertial sensing}
\date{\today}

\begin{abstract}
We present a compact, fibre-coupled interferometer with high sensitivity and a large working range. We propose to use this interferometer as a readout mechanism for future inertial sensors, removing a major limiting noise source, and in precision positioning systems. The interferometer’s peak sensitivity is \(2 \times 10^{-{14}}\)\,m/\(\sqrt{\rm{Hz}}\) at 70\,Hz and \(7 \times 10^{-{11}}\)\,m/\(\sqrt{\rm{Hz}}\) at 10\,mHz. If deployed on a GS-13 geophone, the resulting inertial sensing output will be limited by the suspension thermal noise of the reference mass from 10\,mHz to 2\,Hz. 
\end{abstract}
\maketitle

\section{Introduction}
On the 14th September 2015 the Advanced Laser Interferometer Gravitational-wave Observatory (LIGO) made the first direct detection of gravitational waves \cite{gw150914,gw151226}. To achieve the extraordinary sensitivity required for this discovery, Advanced LIGO uses a complex configuration of suspended mirrors to enhance the signal-to-noise performance of the detector. The mirrors are held at a precise operating point via closed-loop feedback systems to ensure that the laser light is resonant in the various optical cavities in the interferometer. 

In order to reduce the required feedback forces, and associated noise, all core interferometer components are placed on Internal Seismic Isolation (ISI) systems to reduce their inertial and relative motion. The ISIs employ many high-precision inertial and position sensors to reduce the transmission of ground motion \cite{matichardlantz2015}. Additionally, the core optics are mounted inside multi-stage suspension systems that are actively damped using local position sensors \cite{aston2011, strain12}. 

Motivated by the goal of improving local sensing in gravitational-wave detectors, we present a compact interferometer based on the EUCLID and ILIAD sensors developed at Birmingham \cite{Speake05, aston2011, Pena13}. There are two specific applications within LIGO where such a device could be readily employed: as a replacement for the local position sensors in the suspensions, currently shadow-sensors called BOSEMS \cite{Carbone12}; and as a replacement for the coil-magnet readout of Geotech GS-13 geophones. With a focus on the second application, we develop sensitivity requirements to be of interest for LIGO and make an estimate of the potential impact on the observatories.

There exist, however, a large range of other possible applications. Within our narrow focus we include a comparison with past compact interferometers and other LIGO position sensors, and an analysis of the performance of a Watt’s Balance~\cite{bertolini06} with interferometric readout. Wider applications include, but are not limited to, atom interferometers~\cite{Miffre06, Zhou12, Zhou15}, particle accelerators~\cite{Collette12}, and drag-free control of satellites~\cite{Speake05}.

\section{Inertial sensor readout}

The inertial sensors employed by LIGO have internal noises that are substantially higher than the suspension thermal noise limit of their proof masses \cite{Saulson90, Barzilai98}. The readout mechanisms used in high precision inertial sensors are generally either inductive, capacitive, or optical. Capacitance based readouts can achieve high precision (e.g.~\cite{dong11}) but the sensor electrodes must be positioned very close to the target object, limiting their operating range. They also apply significant forces to the object, as well as having a large spatial force derivative (i.e. stiffness), which may be problematic for a suspended mass. This can be partially alleviated by use of multiple electrodes whose contirbution to the force and stiffness can be made to cancel \cite{josselin99}, but the residual effect may still be too great for some applications. Additionally, because the electrodes generally comprise extended plates, the capacitance will depend on some combination of displacement and attitude, directly coupling tilt to displacement. The drive signal of capacitance measurements may also pollute their environment with audio frequency electric fields, which are undesirable in sensitive experiments such as GW detectors \cite{bertolini06}.

Inductive sensors suffer from many of the same technical issues as capacitive sensors, including the trade between sensitivity and range and cross-coupling, but they are even more sensitive to EM interference~\cite{Tariq02, Matichard15}. A final class of readout scheme, which should be considered separately from classical coil-driver inductance measurements, employs superconducting inductance measurements such as SQUIDs \cite{strayer03}. These can achieve very high sensitivity, but their cryogenic nature clearly makes them expensive and impractical for many applications.

Using interferometers to measure the proof mass position has the potential to remove some of the existing limitations in readout and actuation noise, while circumventing the technical challenges inherent in capacitative, inductive, and superconducting sensors. Other groups \cite{Zumberge04, Zumberge10, Watchi16, Arp13, Venkateswara15} have had success in improving the performance of inertial sensors using optical readout to both commercial and custom mechanics. We propose to extend the state-of-the-art by combining interferometric readout with commercial inertial sensor mechanics, improving sensitivity below the noise floor of the best force-feedback seismometers, such as the Trillium T240~\cite{t240spec}. 

\section{Sensitivity requirements}
At the LIGO detector sites the ground motion at 10\,Hz is approximately 10 orders of magnitude larger than measured gravitational-wave signals. The use of complex multi-stage passive and active isolation systems attenuates input motion below other noise sources at frequencies above 10\,Hz \cite{AdvancedLIGO15}. Seismic noise at frequencies below 1\,Hz lies outside the sensitive band of the interferometer. Nevertheless, ground motion at these frequencies, where active feedback provides most of the isolation, can still increase the RMS motion of the interferometer mirrors enough to prevent operation. The primary contributions to residual motion between the optics (excluding earthquakes) comes from the secondary micro-seismic peak (typically between 0.15 and 0.35\,Hz) and the coupling between tilt and translation (typically below 0.1\,Hz) \cite{Lantz09}. 

It is difficult to predict the effect of new instruments on LIGO, the control systems and behaviour of the instrument is extremely complex. However, during the first observation runs, Advanced LIGO was unable to operate for approximately 18\% of the time due to elevated wind and microseismic motion~\cite{mowlowry18}. By reducing the RMS motion of the isolation platforms, the interferometer should be able to operate during a wider range of environmental conditions. Moreover, due to the implementation of phasemeter readout (sometimes called fringe counting), our interferometric sensors have a larger working range than both the GS-13's and T240's employed at LIGO. The extra range and improved low-frequency sensitivity may improve the detectors' ability to stay `locked' during small or remote earthquakes by suppressing only the differential inertial motion.

The control band for LIGO's active inertial isolation for the ISIs is approximately 100\,mHz to 30\,Hz \cite{Matichard15}. At low frequencies the noise on the inertial sensing output increases as 1/$f^2$ and as such, the inertial signal is substituted with displacement sensors effectively locking the isolated platforms to to the ground below approximately 30\,mHz. However due to the constraints of causal filtering, the inertial sensors must perform well down to 10\,mHz to avoid injecting sensor-noise or tilt-coupling. Performance requirements between 1 and 10\,Hz mean that the unity gain frequency must be about 30\,Hz, and as such, good inertial sensor performance (in both sensitivity and phase response) is needed up to 100\,Hz. Beyond 100\,Hz it is possible to rapidly reduce the loop gain and the sensitivity requirements are subsequently relaxed. For these reasons, to be of interest for Advanced LIGO (and other gravitational-wave detectors), any new inertial sensor should have sensitivity at least equal to state-of-the-art inertial sensors between 10\,mHz and 100\,Hz.

Further improvements to the detector's performance can be made by increasing the sensitivity of the BOSEM displacement sensors placed on LIGO's quadruple suspensions \cite{Carbone12}. Due to the noise of the BOSEMs, local feedback forces can only be applied to the uppermost suspended mass, and even then the control filters have strict requirements imposed by the need to prevent sensor noise from spoiling the detector sensitivity at 10\,Hz \cite{strain12}. Interferometric displacement sensors would allow for improved damping of the top mass of the quad suspension system, as well as opening the possibility for local-damping on lower stages, reducing both vibration transmission and settling time. To apply significant damping using a sensor at the Upper-Intermediate Mass \cite{Aston12} (the second stage of the `quad' suspension from the ground), the noise of the sensor at 10\,Hz should be of order 100 times smaller to exceed the increase in mechanical transmissibility at this frequency, and our measurements here more than satisfy this criterion.

\section{Readout Scheme and Optical Layout}
A standard two-beam interferometer has an operating range that is typically less than a quarter of a wavelength of path-length difference. To increase both the dynamic range and the operating range, without using actuators or modulation schemes, we employ a Homodyne Quadrature Interferometer (HoQI) that can measure two nearly orthogonal quadratures of the interferometer output. In this case, we use a Mach-Zender interferometer with two independent recombination beamsplitters. A polarisation scheme is employed to generate the required differential phase shift \cite{Bouricius70}.

\begin{figure}[!ht]
	\centering
	\includegraphics[width=0.45\textwidth]{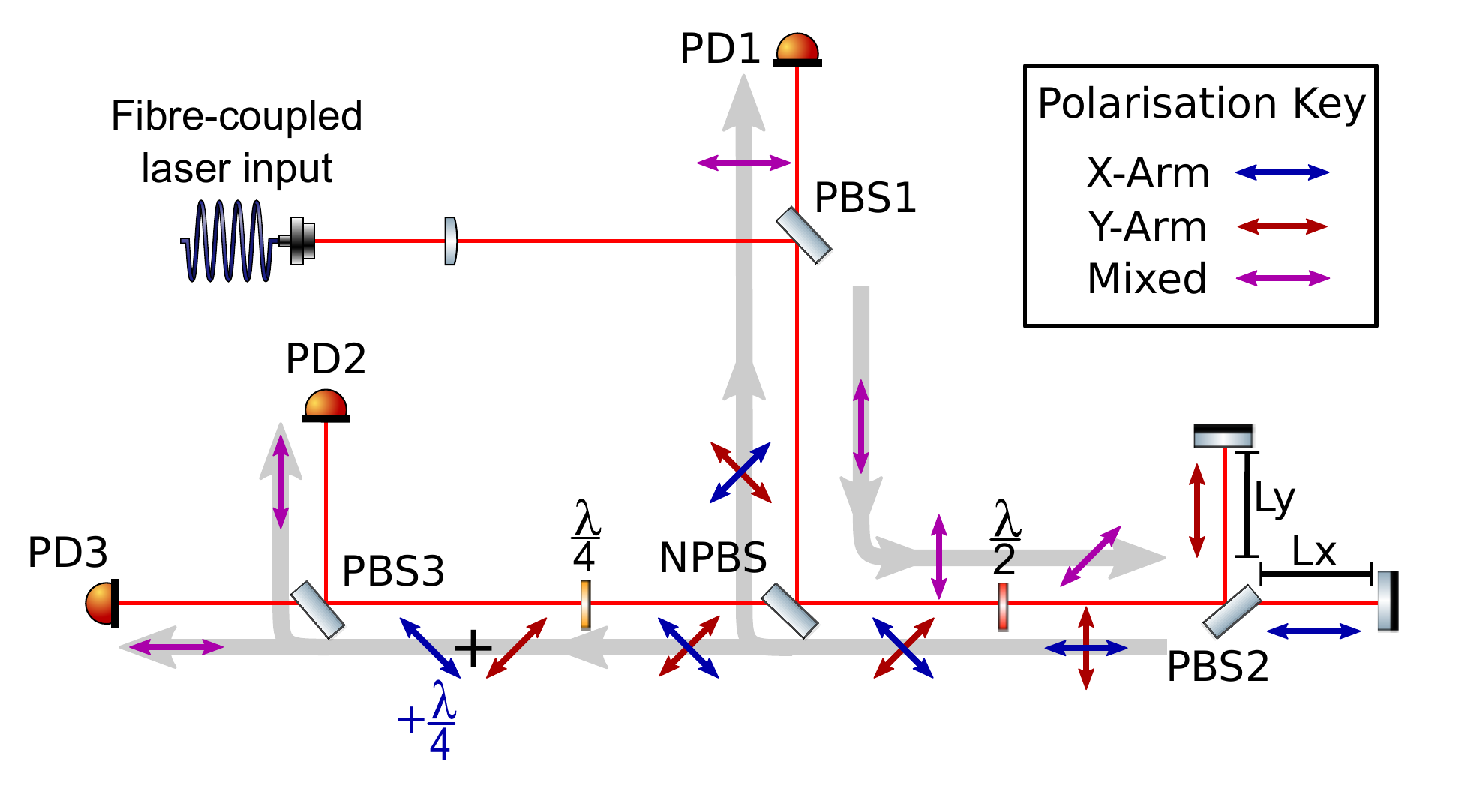}
	\caption{The optical layout of HoQI. Orthogonal polarisation states are used to track the length difference between Lx and Ly over multiple optical fringes. The input beam is split at polarising beamsplitter PBS2 and interferometrically recombined at PBS1 and PBS3, producing signals proportional to the sine, cosine, and minus cosine of the optical phase difference. Grey arrows indicate the direction of propagation.}
	\label{fig:HoQIOptics}
\end{figure}

The optical path of HoQI is shown schematically in Fig.~\ref{fig:HoQIOptics}. Compared with EUCLID and ILIAD, it is significantly simpler. The number of birefringent elements, which increase noise and non-linearity, has been substantially reduced, and there are no longer waveplates in the `arms' of the interferometer. The tilt-compensation system, developed to reduce tilt-to-length coupling and increase the angular operating range of the instrument~\cite{Pena11}, was also removed. The double-pass nature of the `cat's eye' system resulted in parasitic interferometers with a relatively large arm-length mismatch, which in turn couples frequency changes in the laser into measurement error. 

To further reduce frequency noise coupling, we use a narrow-linewidth 1064\,nm solid-state Innolight Mephisto 500NE laser (1\,kHz linewidth for 0.1\,s averaging period) in place of the VCSEL diode laser, and we carefully match the arm lengths. 

During our first tests of HoQI, we use carefully-aligned high-stability steering mirrors. To interrogate external targets, such as an inertial sensor reference mass, we will need to increase the angular operating range. At present we intend to use a double-pass lens, placing a small focus on the target mirror. This will increase the divergence angle of the beam (up to $\sim$10\,mrad), making us less susceptible to misalignment. For cases where larger operating ranges are required, a corner cube can be placed on the remote optic, and a large beam-size ($\sim$2\,mm) used within the interferometer, allowing operation over (at least) several degrees. 

Assuming the target mirror remains aligned the operating range of HoQI is only limited by the fringe visibility degradation due to spot-size changes, and it is more than 10\,mm for this configuration.

The laser light is fibre-coupled to the interferometer by a 2\,m single-mode polarisation maintaining fibre with an input power of 10\,mW. The first Polarising BeamSplitter, PBS1, ensures there is a clean input polarisation state. PBS2 splits the input beam into two orthogonally polarised beams, one for each arm. These beams are recombined at PBS2 and co-propagate without interfering. The beam is divided, again without interference, at the Non-Polarising BeamSplitter (NPBS).  The quarter-wave plate before PBS3 then adds an additional phase shift of 90\,degrees to the light from one of the arms such that when the beams interfere at PBS1 and PBS3, the resulting intensity fluctuations are 90 degrees out of phase. 

The power measured on the photodiodes is given by the following equations,
\begin{eqnarray}
\rm{PD1} & = & \frac{P_{in}}{8}(1+a\sin(\phi_{opt})) \label{eqn:PD1},\\
\rm{PD2} & = & \frac{P_{in}}{8}(1+a\cos(\phi_{opt}))\label{eqn:PD2}, \\
\rm{PD3} & = & \frac{P_{in}}{8}(1-a\cos(\phi_{opt})) \label{eqn:PD3}, \\
\rm{PD1 -PD2} & = & \frac{\sqrt{2}aP_{in}}{8}\sin(\phi -\frac{\pi}{4}) \label{eqn:PD1-PD2}, \\
\rm{PD1 -PD3} & = & \frac{\sqrt{2}aP_{in}}{8}\sin(\phi +\frac{\pi}{4}) \label{eqn:PD3-PD2},
\end{eqnarray} where \(P_{in}\) represents the input power, \textit{a} is the fringe visibility and \(\phi_{opt}\) represents the differential optical phase and is defined as \(\phi_{opt}=\frac{4\pi(L_x-L_y)}{\lambda}\). Equations \ref{eqn:PD1-PD2} and \ref{eqn:PD3-PD2} show how these signals can be combined to provide substantial common-mode rejection of laser intensity noise by reducing the dependence on both the input power and the fringe visibility. 

Unwrapping the 4-quadrant arctangent of equations \ref{eqn:PD1-PD2} and \ref{eqn:PD3-PD2} returns the optical phase. To achieve high resolution, each photodiode signal is digitised with a high dynamic range, 18-bit ADC and the arctangent is performed using a cordic engine implemented on an FPGA. The analogue front-end and digital processing use an electronics module developed for the EUCLID and ILIAD interferometers \cite{Speake05}, which have exceptionally low input-referred noise at low-frequencies and a proven signal processing chain. The displacement-equivalent noise of the readout electronics is shown in Fig.\ref{fig:HoQIASD}, and it is what enables the high precision reported here.
\begin{figure}[!ht]
	\centering
	\includegraphics[width=0.45\textwidth]{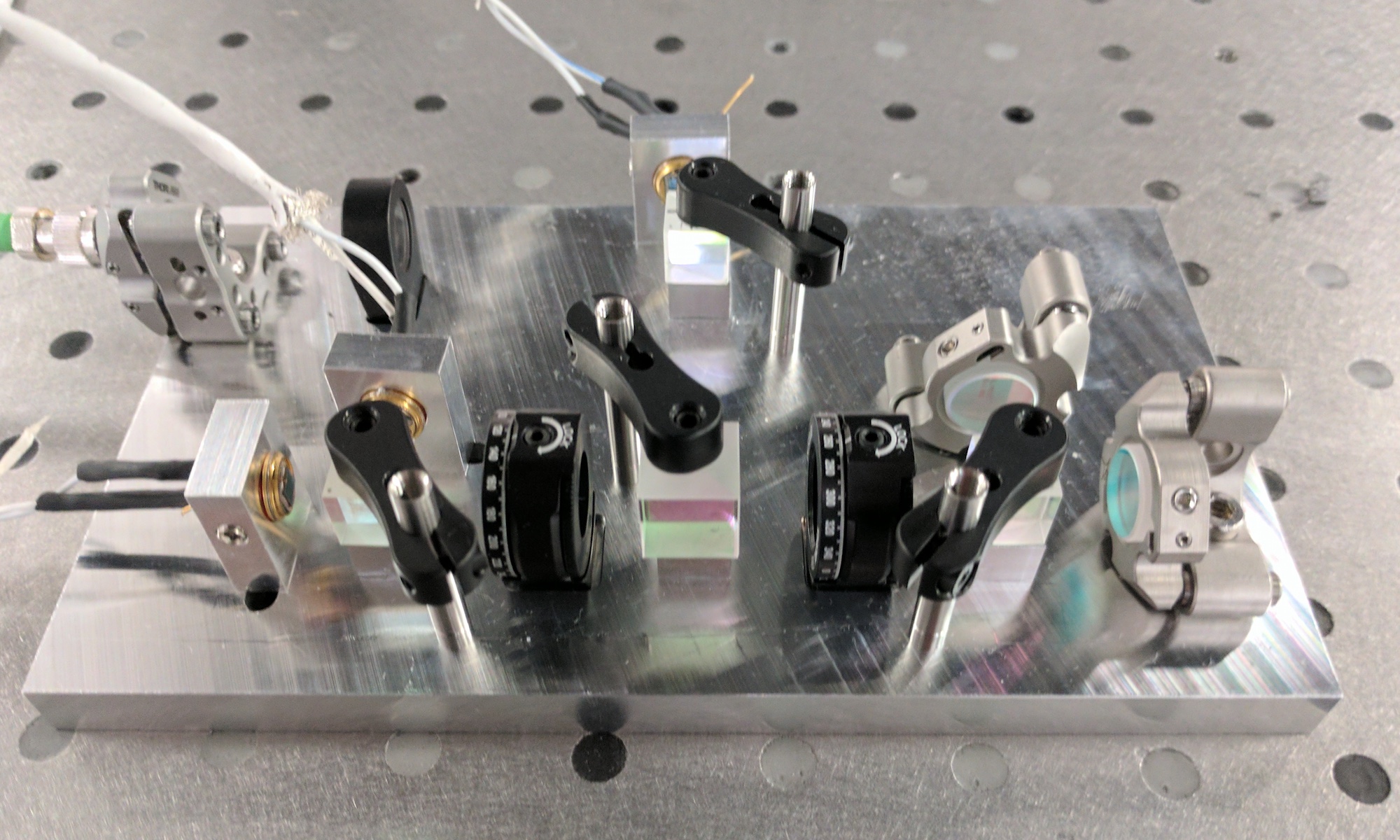}
	\caption{The prototype version of HoQI, the base plate is $170 \times 100\,$mm with 10\,mm gaps between components.} 
	\label{aHoQIimage}
\end{figure}

\section{Results}

To investigate the sensitivity limits of HoQI we reduced optical and mechanical noise where possible. The largest anticipated sources of noise were: mechanical vibration, thermal expansion and gradients, birefringence noise, frequency noise, and electronic noise. All optics were rigidly mounted close together on an aluminium baseplate with a relatively large thermal mass, seen in Fig. \ref{aHoQIimage}, resulting in large common-mode rejection of mechanical noise and reducing thermal gradients. 

Birefringence fluctuations between the non-polarising beamsplitter and the recombination polarising beamsplitters are indistinguishable from arm-length changes. Since the beams are well aligned and co-propagate, the dominant effect is expected to come from quarter-wave plate, and a high quality zero-order waveplate was used to reduce this. Alignment fluctuations on the photodiodes cause uncorrelated fluctuations in the photocurrent due to inhomogeneities in the quantum efficiency across the surface of the photodiode \cite{Kwee09}. The single-mode fibre strips away pointing fluctuations, and the output mode is mechanically fixed to the baseplate by the fibre output collimator.

\begin{figure}[!ht]
	\centering
	\includegraphics[width=0.49\textwidth]{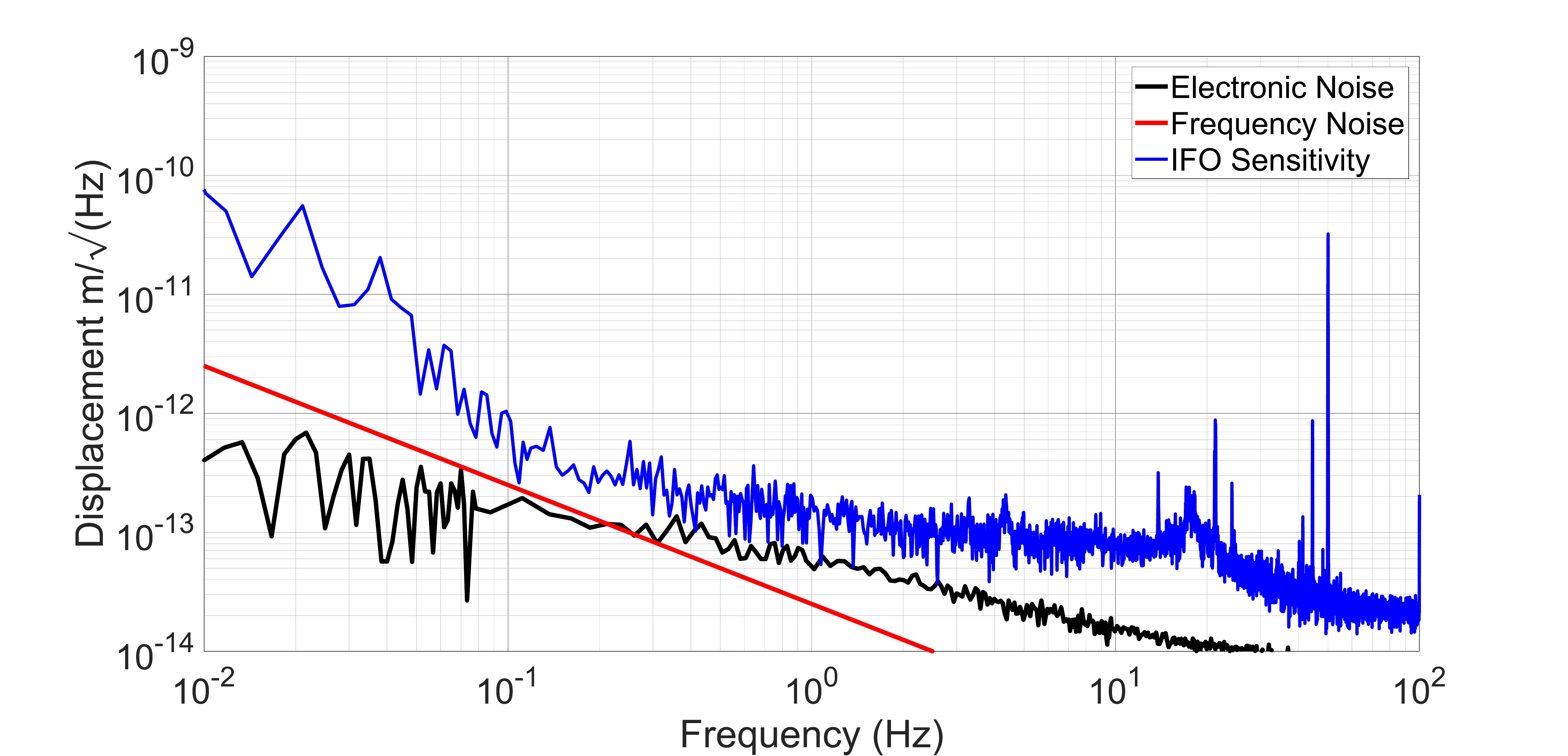}  
	\caption{Sensitivity of the fibre-coupled prototype HoQI showing the interferometer signal (blue), the measured readout noise (black), and an estimate of the frequency noise that couples into the interferometer (red)}
	\label{fig:HoQIASD}
\end{figure}

Frequency noise coupling was measured and minimised by modulating the laser frequency and adjusting the macroscopic arm-length difference to minimise the coupling to differential optical phase. The length was precisely tuned using the alignment screws on the `end' mirrors, with a resolution of a few microns, but the coupling was much larger than predicted. This is attributed to interference from stray light. The residual coupling can be quantified by an effective arm-length mismatch of 0.7\,mm. Assuming laser frequency fluctuations of $10^4 \times [\frac{1}{f}]$\,Hz/$\sqrt{\rm{Hz}}$, we predict the red curve shown in Fig. \ref{fig:HoQIASD}. 

The electronic noise (the black curve in \ref{fig:HoQIASD}) is measured by replacing the photodiode inputs with a constant current using a resistor connected to a bias voltage. The resistor values are such that the 3 input currents simulate a specific optical phase for the three photodiodes. 

The baseplate was placed on rubber `feet' on an optical bench and sampled at 20\,kHz over a 10 hour period. Fig.\ref{fig:HoQIASD} shows the amplitude spectral density of the measurement over a ten minute segment of this data. The interferometer reaches a peak sensitivity of \(2 \times 10^{-{14}}\)\,m/\(\sqrt{\rm{Hz}}\) at 70\,Hz. At 10\,mHz a sensitivity of \(7 \times 10^{-{11}}\)\,m/\(\sqrt{\rm{Hz}}\) is achieved.

The total sensitivity is probably limited by electronic noise at frequencies near 0.5\,Hz. Below this, the limiting factor is assumed to be a combination of air currents, temperature fluctuations, and frequency noise.  Above 1\,Hz, the sources of noise are less well understood except for the peak near 18\,Hz, that is caused by mechanical vibration of the optical table, and the large peak at 50\,Hz, caused by pickup in the unshielded photodiode cables.

\begin{figure}[!ht]
	\centering
	\includegraphics[width=0.49\textwidth]{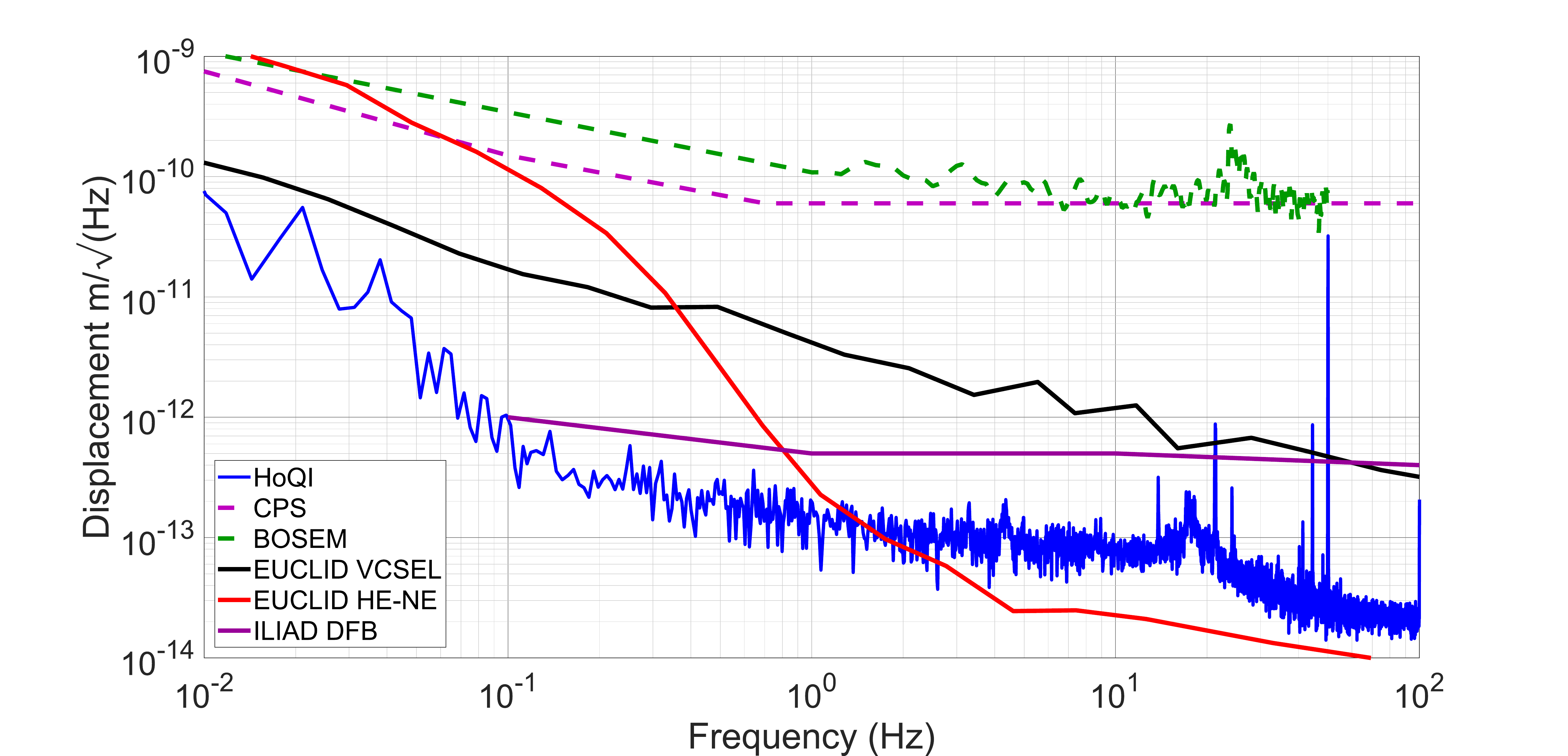}		
	\caption{HoQI (blue) compared with other precision displacement sensors including: previous interferometers developed at Birmingham, ILIAD (purple)~\cite{Pena13} and EUCLID with both an external HE-NE laser (red) and with its integrated VCSEL laser (black)~\cite{aston2011}; and with devices used at LIGO, the 0.25\,mm range Capacitative Position Sensor (CPS, dashed purple), BOSEM (dashed green). The CPS and low-frequency BOSEM curves are stick-figure fits to noise spectra from multiple devices.}
	\label{fig:aHoQIvPosSensor}
\end{figure}

Fig.\ref{fig:aHoQIvPosSensor} compares the sensitivity of HoQI with the Capacitive Position Sensors (CPS), which are employed on the first stage of LIGO's Internal Seismic Isolation system (ISI). In the frequency band of interest they offer 250 times lower noise at 100\,mHz and 1000 times lower noise at 10\,Hz. When compared with the BOSEMs, the improvement is more substantial: HoQI has a factor of 500 lower noise at 100\,mHz and 1000 times lower noise at 10\,Hz. 

\begin{figure}[!ht]
	\centering
	\includegraphics[width=0.49\textwidth]{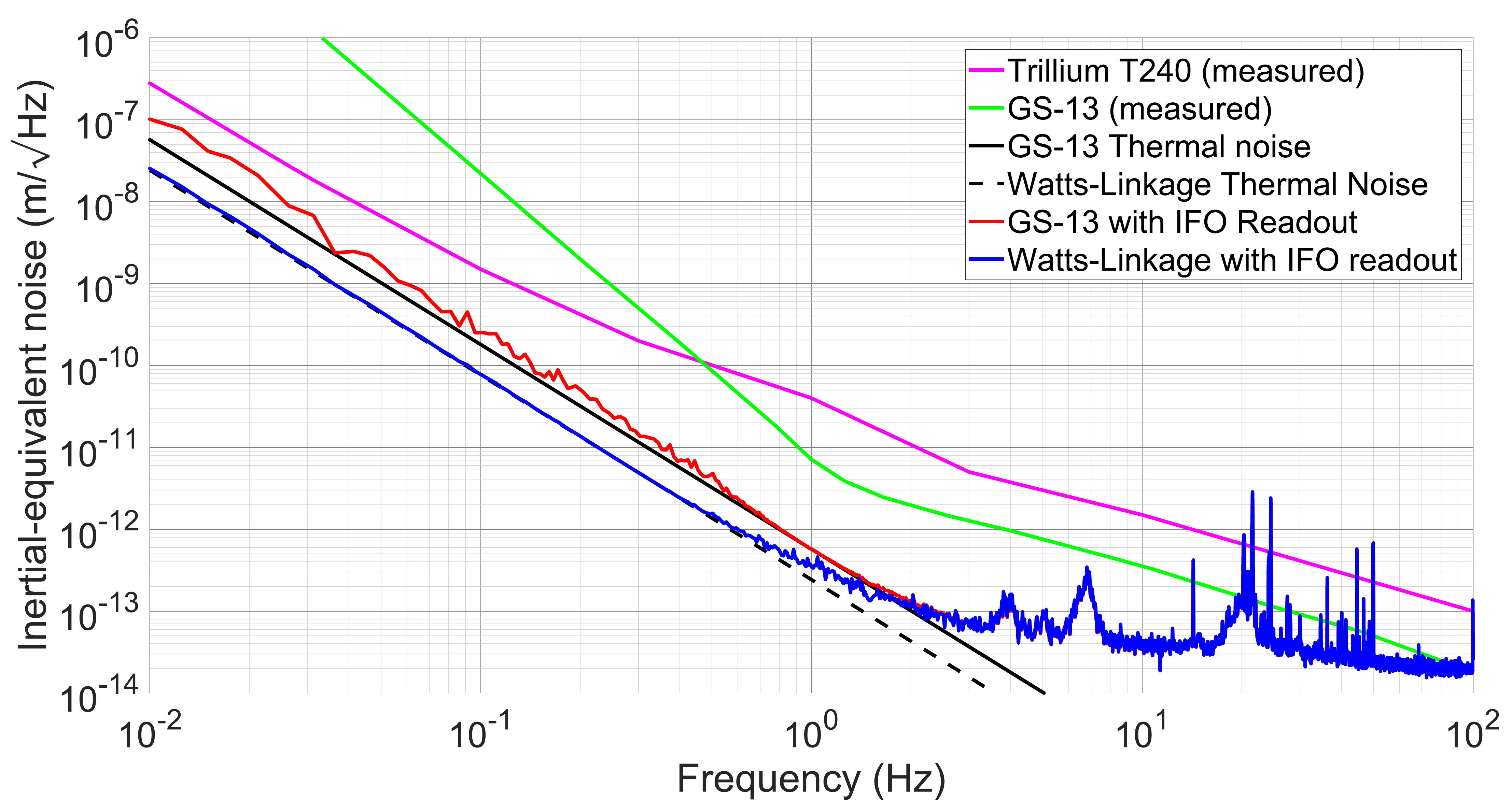}
	\caption{The sensitivity of HoQI projected onto a GS-13 (red) and a Watt's linkage (blue) is compared with a GS-13 using conventional readout (green), and a Trillium T-240 force-feedback seismometer (magenta). The (calculated) suspension thermal noise of the GS-13 (black) \cite{GS13modify} and Watt's linkage (dashed black) are also shown.}
	\label{fig:HoQIPlantASD}
\end{figure}

In order to compare HoQI's readout noise with existing inertial sensors, we multiply the interferometer sensitivity curve by the inertial-sensing transfer function of both a GS-13 and a Watt's linkage similar to those employed at the Virgo gravitational-wave detector \cite{bertolini06}. The result of this is shown in Fig.~\ref{fig:HoQIPlantASD}. This readout-noise is then summed in quadrature with the estimated suspension thermal noise for each sensor. The mechanical thermal noise is given by,
\begin{eqnarray}
F_{\rm th}(\omega)  = \sqrt{\left( \frac{4 k_B T R(\omega)}{Q} \right)},
\end{eqnarray}
where $F_{\rm th}(\omega)$ is the amplitude spectral density of the force due to thermal noise, $T$ is the temperature, $Q$ is the quality factor and $R(\omega)$ is the mechanical resistance (the real part of the mechanical impedance) \cite{Saulson90}. For a simple mass-spring system with mass $m$, resonant frequency $\omega_0$, the mechanical resistance is given by,
\begin{eqnarray}
R(\omega) = \frac{m\omega_0^2}{Q\omega}
\end{eqnarray}  

The GS-13 is assumed to have a 5\,kg proof-mass, a resonant frequency of 1\,Hz, and a (structural-damping) quality factor of 40. The Watt's linkage, with its low mechanical-dissipation and resonant frequency, has lower thermal noise (everywhere) and lower readout noise below 1\,Hz. For the suspension thermal noise calculation we assume a proof-mass of 1\,kg, a resonant frequency of 0.3\,Hz, and a (structural-damping) quality factor of 100. 

The noise projections are compared with the self-noise floors of the GS-13 (using it's conventional coil magnet readout) and a Trillium T240, both as measured at LIGO. We find that between 0.01 and 2\,Hz the suspension thermal noise of the GS-13 would limit the resolution of an future optically readout inertial sensor, based on GS-13 mechanics. To fully exploit the sensitivity of of the interferometer presented in this paper mechanics with a lower suspension thermal noise would have to be evaluated. Increasing the structural Q of the spring reduces this thermal noise, and an improvement in the resolution between the optically readout GS-13 and Watts linkage can be seen. Despite the thermal noise limitation, using HoQI to interrogate a GS-13 could increase the sensitivity by a factor of 100 at 100\,mHz and would improve it at all frequencies up to 100\,Hz.

\section{Conclusion}

We have presented a new compact interferometer that employs a homodyne phasemeter, HoQI. This combines an existing architecture with a low-noise laser and readout system to achieve excellent noise performance from 10\,mHz to 100\,Hz. The sensitivity is substantially better than existing displacement sensors at LIGO, sufficient to improve the performance of suspension damping systems. If used as part of an inertial sensor using existing mechanics, it could reduce the self-noise across a wide range of frequencies, down to the suspension thermal noise of the mechanical springs.

\begin{acknowledgments}

We thank John Bryant for technical support. This project has received funding from the European Union's Horizon 2020 research and innovation programme under the Marie Sk\l{}odowska-Curie grant agreement Number 701264.
\end{acknowledgments}

\bibliographystyle{apsrev}

\end{document}